\documentclass[11pt]{article}
\RequirePackage{amsthm,amsmath}
\RequirePackage{natbib}
\RequirePackage[colorlinks,citecolor=blue,urlcolor=blue]{hyperref}
\usepackage{bm}
\usepackage{amssymb,amsthm,amsmath}  
\usepackage{array,booktabs}
\usepackage{multirow}
\usepackage{graphicx}
\usepackage{color,xcolor}
\usepackage{comment}
\usepackage{color,xcolor}
\usepackage{tabu}
\usepackage{pdflscape}
\usepackage{url}

\title{Landscape allocation: stochastic generators and statistical inference}

\author{Patrizia Zamberletti, Julien Papa\"{i}x, Edith Gabriel, Thomas Opitz \\
	Biostatistics and Spatial Processes, INRAE, Avignon, France
}

\begin{document}
\maketitle






\begin{center}
	\begin{minipage}{0.8\linewidth}
		{\textbf{Abstract. }}

In agricultural landscapes, the composition and spatial configuration of cultivated and semi-natural elements strongly impact species  dynamics, their interactions and habitat connectivity. To allow for landscape structural analysis and scenario generation, we here develop statistical tools for real landscapes composed of geometric elements including 2D patches but also 1D  linear elements such as hedges.  We design generative stochastic models that combine a multiplex network representation and Gibbs energy terms to characterize the distributional behavior of landscape descriptors for land-use categories. We implement Metropolis-Hastings for this new class of models to sample agricultural scenarios  featuring parameter-controlled spatial and temporal patterns (e.g., geometry, connectivity, crop-rotation). Pseudolikelihood-based inference allows  studying the relevance of model components in real landscapes through statistical and functional validation, the latter achieved by comparing commonly used landscape metrics between observed and simulated landscapes. Models fitted  to subregions of the Lower Durance Valley (France) indicate strong deviation from random allocation, and they realistically capture small-scale landscape patterns. In summary, our  approach of statistical modeling improves the understanding of structural and functional aspects of agro-ecosystems, and it enables simulation-based theoretical analysis of how landscape patterns shape biological and ecological processes.

\end{minipage}
\end{center}

\newpage 



\section{Introduction}

Agroecosystems  are the basis for food production and other ecosystem services  such as biodiversity, pollination and pest control \citep{Power2010, Foresight2011}. Landscape heterogeneity plays an important role for many agroecological processes and can be expressed  through landscape \emph{configuration}, referring to the size, shape, and spatial-temporal arrangement of land-use patches, and through landscape \emph{composition}, referring to the number and proportion of land-use types \citep{Martin2019}.  
Generative models are widely applied in landscape ecology for simulating   virtual landscapes (i.e., a mosaic of fields having shapes and properties that vary in space and time including or not  biotic and abiotic relationships), differing in configuration and composition, to systematically study the effects and impacts of landscape heterogeneity on ecosystem processes; see the reviews of \citet{Poggi2018,Langhammer2019}. Such models should allow generating a high number of virtual but structurally realistic maps of land cover \citep{Gardner1999, Saura2000, Gardner2007, Sciaini2018}, and often parameters related to landscape features such as the percentage of land-cover, the habitat fragmentation, or  spatial autocorrelation \citep{Langhammer2019} can be controlled. 
In this paper, we  focus on agricultural landscapes, and we consider neutral landscape models where the model does not directly interact with biotic or abiotic processes \citep{Gardner1987, With1997}. 

Models use either a vector-based or a raster-based representation, with the majority of models in the literature being of raster type. The raster approach is  particularly useful for modelling gradual landscape dynamics and continuous processes. However, agricultural landscapes are strongly characterized by polygon-shaped patches and piecewise linear corridors following polygon boundaries, such that vector approaches seem preferable \citep{Gaucherel2006a, Gaucherel2006b, LeBer2009, Papaix2014, Inkoom2017, Langhammer2019}. In particular, fringe structures such as hedgerows, roads or ditches aligned along polygon boundaries have an important impact on many agroecological processes. 
In a vector-based framework, \citet{Gaucherel2006a, Gaucherel2006b}  use models based on Gibbs energy terms to control certain pairwise interactions between landscape elements with the aim of simulating patches and certain fringe structures. \cite{Papaix2014} develop a landscape generator without fringe structures that generates the landscape mosaic with two types of fields based on the T-tessellation algorithm of \citet{Kieu2013}. However, existing modeling frameworks lack tools for parameter inference and model validation. Validation procedures are usually solely based on testing whether simulated landscapes are able to reproduce realistic landscape features by comparing landscape metrics (e.g., from the FRAGSTAT library \citet{Mcgarigal1995}). Sometimes, such metrics are directly used within simulation algorithms to enforce convergence towards target values \citep{Langhammer2019}. 

In this paper, we advocate to turn away from the raster paradigm when modeling agricultural landscapes. Instead, vector-based approaches are independent of the grid resolution and give better control over small-surface elements, and they provide a sparser and more functional representation of patchy geometric structures without continuous gradients. The approach that we develop is geared towards flexible and realistic parametric stochastic modeling of fringe structures, such as hedgerow networks. Based on a network representation of interactions among landscape elements, we construct Gibbs energies pertaining to exponential family models, which provides a natural distributional framework for controlling landscape descriptors.  

We assume that the polygon structure of patches in a subset of planar space $\mathbb{R}^2$ is given, i.e., a tessellation of space serves as fixed support of the model. It can be obtained by preprocessing a real landscape, or we can use simulations of a parametric tessellation model to generate realistic features.
We model the stochastic land-use allocation mechanism of patches and linear elements by  assigning categories to the polygons and their edges, where dynamic structures such as crop rotation are possible. 
The composition of landscape is expressed through the proportions of categories (e.g., numbers, relative lengths or surface areas of objects), while configuration relates to  the spatial-temporal arrangement of categories, such as clustering or repulsiveness.

An overarching goal  is to generate visually realistic landscapes, and we develop the following methodological novelties:  
i) mathematical representation of landscape composition and configuration through a multilayer network;
ii) generative stochastic parametric models coupling land-use allocation of patches and linear elements;
iii) simulation of such models using Markov-Chain Monte-Carlo (MCMC);
iv) statistical inference using  real landscape data;
v) validation over real landscape data by comparing metrics for vector and raster representations between real and simulated landscapes. 
Our approach can handle relatively large landscapes by capitalizing on low computational requirements thanks to vector-based representations and to sparse-matrix structures for interactions. 

In the remainder of the paper, Section~\ref{sec:data} presents real landscape data and preprocessing steps for an agricultural region in southeastern France, for which previous studies have highlighted a key role of agricultural practices and hedgerow configuration  for biodiversity and pest control \citep{Ricci2009, Maalouly2013, Lefebvre2016}.
In Section \ref{method_Stochastic_modeling}, we propose the mathematical representation, modeling and simulation of landscapes. 
Tools for statistical inference and validation are developed in Section~\ref{method_Stat_inf}. In Section \ref{method_BVD}, we apply the developed framework to the above data, and we discuss how the goodness-of-fit and the generation of realistic landscape metrics is influenced by the choice of the descriptors in the model. A discussion in Section \ref{Concl} concludes the paper. Supplementary material contains details on the simulation algorithm and additional estimation results.

\section{Landscape data}\label{sec:data}

\begin{figure}[h!]
\centering
\includegraphics[scale=0.4]{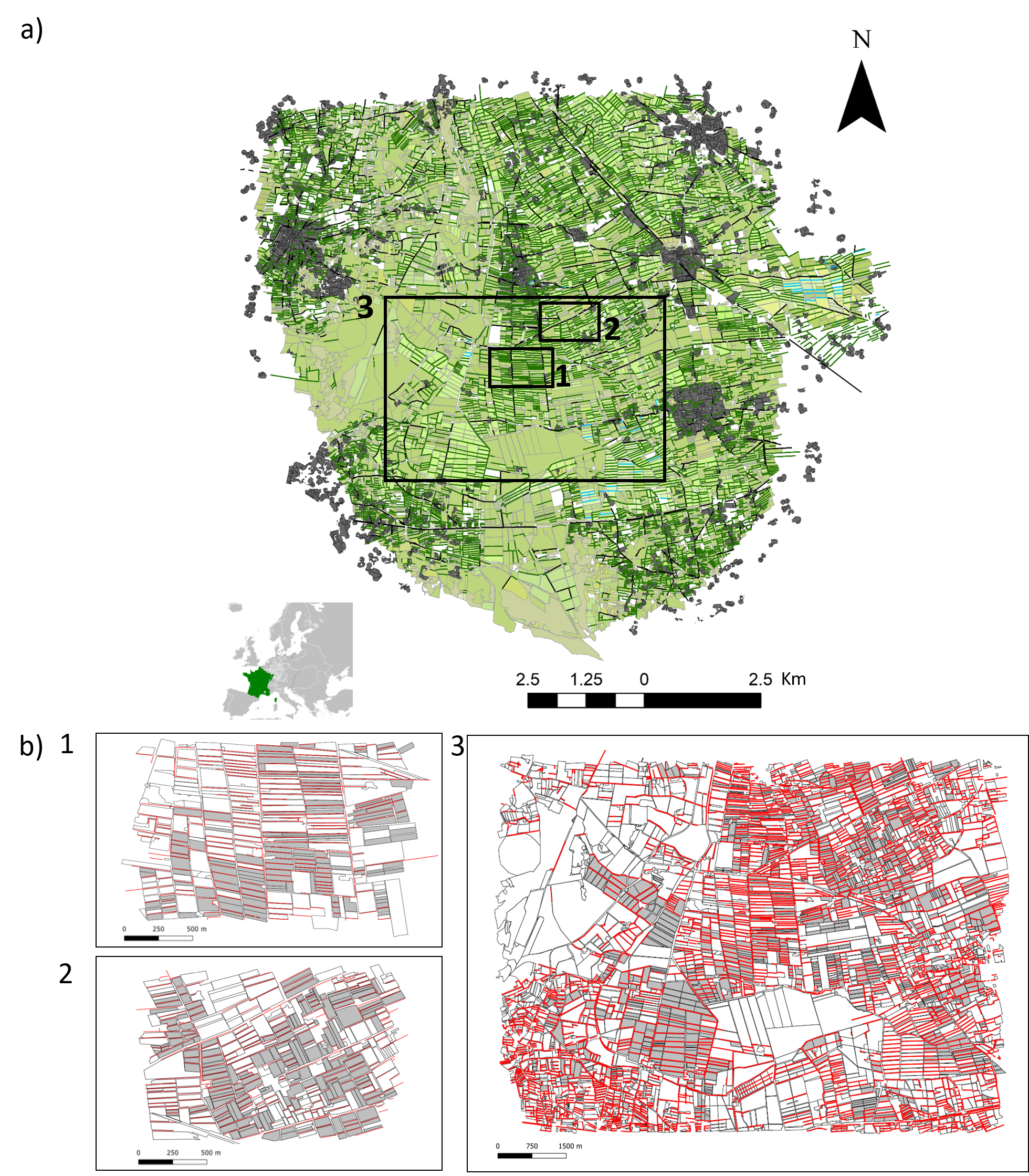}
\caption{Lower Durance Valley study area. a) Full area with $3$ subdomains. b) Subdomains: 1) \textit{small region D1}; 2) \textit{small region D2}; and, 3) \textit{large region D3}.}
\label{fig:Fig2}
\end{figure}

Real data for agricultural landscapes are based on remote sensing images, digital land registers, land cover data bases such as CORINE \citep{Buttner2006}, and field data. Often, manual annotation steps are necessary to complete and clean data. 
We study the Lower Durance Valley in southeastern France depicted in Figure \ref{fig:Fig2}a, stretching over $163.06 km^2$ and mainly characterized by agricultural activity ($87$ \%) and urbanized areas, with main cultures of open area (46\%) and  apple/pear orchards (24\%).

Data are based on manual digitalization with the ArcView software using an official French  database of  aerial photographs (BD ORTHO, IGN, 2004, 0.5m resolution, updated with field monitoring in 2009).

The region totals $1145.94 km$ of hedgerows, which we will represent as linear segments, with average length of $104.65m$. We distinguish East-West oriented wind-break hedges to break the strong Mistral winds (83\%) from roadside hedges. 

For the data application in this paper, we select three  subdomains with contrasting properties and dimensions:  region 1  in Figure \ref{fig:Fig2}b1 is relatively small and we refer to it as D1; region 2 in Figure \ref{fig:Fig2}b2 has the same surface area and is called D2; and region 3 in Figure \ref{fig:Fig2}b3 delimits a much larger domain including D1 and D2, which we denote by D3. Table \ref{tab:Regions} summarizes details.

\begin{table}[h!]
\caption{Summary of selected subregions of Lower Durance Valley study area; see Figure \ref{fig:Fig2}.}
\label{tab:Regions}
\centering
\begin{tabular}{lccc}
                                           & \multicolumn{3}{c}{Subregion} \\
                                           & D1        & D2        & D3        \\ \hline
\multicolumn{1}{l|}{Area ($km^2$)}            & 3.37     & 2.3      & 41.13    \\
\multicolumn{1}{l|}{$\%$ of Semi-natural}    & 73       & 50       & 76       \\
\multicolumn{1}{l|}{$\%$ of Crop}            & 27       & 50       & 24       \\
\multicolumn{1}{l|}{Hedgerows ($km$)}        & 44.64    & 33.61    & 386.36   \\
\multicolumn{1}{l|}{No. of patches}        & 368      & 468      & 4379     \\
\multicolumn{1}{l|}{No. of linear segments} & 1105     & 1405     & 12517   
\end{tabular}
\end{table}

\begin{figure}
\centering
\includegraphics[scale=0.5]{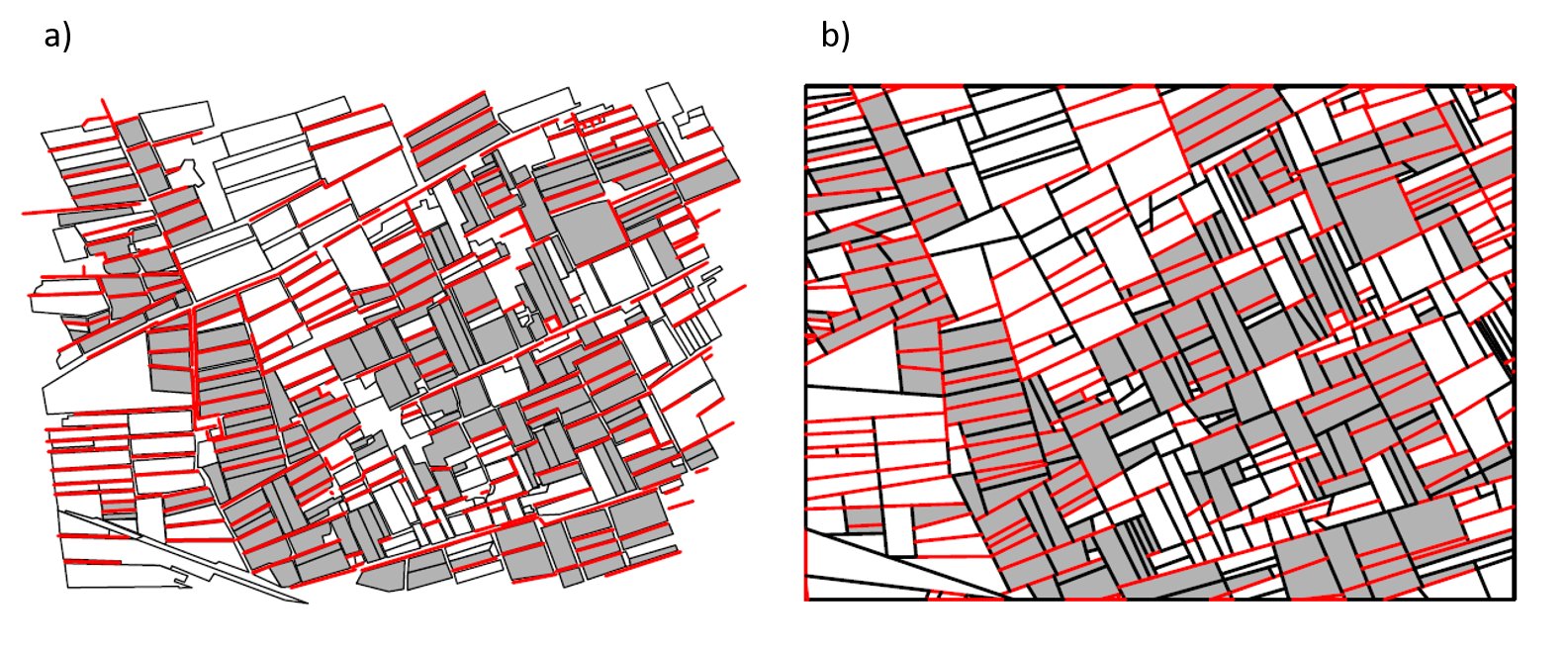}
\caption{Side-by-side representation of the original digitalized shapefile (a) and the landscape tessellation after preprocessing (b) for the \textit{small region D2}.}
\label{fig:D2}
\end{figure}

We use a simplified representation of the landscape as a tessellation of $2D$ space with polygon-shaped cells. Linear segments (e.g., hedgerows) correspond to polygon edges. To achieve polygon shape for patches defining a continuous cover of space, and to align hedgerows with  polygon edges, we preprocess the landscape towards a polygon tessellation of 2D space \citep{Okabe1992}. For this purpose, we minimize a heuristic loss criterion measuring the distance between original and transformed landscape. Figure \ref{fig:D2} illustrates that landscape modifications during preprocessing for domain D2 are mostly minor. In our models, the tessellation will be considered as a fixed support for linear element attribution and crop rotation. Tessellation simulation algorithms for agricultural landscapes \citep{Kieu2013, Papaix2014, Poggi2018} also enable the  generation of new synthetic but realistic supports for our models.

\section{ Stochastic modeling and simulation of landscape allocation} \label{method_Stochastic_modeling}

\subsection{Mathematical landscape representation} \label{method_obj}
We propose to represent a landscape as a collection $\mathcal{O}=\{o_1,\ldots,o_n\}$  of $n$ geometric objects as follows, 
\begin{equation}
   o_i=(z_i,x_i),\quad x_i\in \mathcal{X}_i=\{0,1,\ldots,\ell_i-1\},  \quad  i=1,\ldots,n,
\end{equation}
where each element is composed of two sets of data $z_i$ and $x_i$. The information in  $\bm z=(z_1,\ldots,z_n)$ is considered as being fixed, while the data $x_i$ contains information on  the $i$th geometric element that we aim to model. The vector $\bm x=(x_1,\ldots,x_n)$ represents categories that we allocate to the geometric elements in the landscape, such as the land-use types. We suppose that $x_i\in \mathcal{X}_i$ with a finite space  $\mathcal{X}_i$ of $\ell_i\geq 1$ possible categories for the $i$th element. 
The objects $o_i=(z_i,x_i)$ could represent different types, such as polygons (i.e., habitat patches) or linear segments (i.e., linear landscape elements), see Figure \ref{fig:Fig1}a.
For polygon objects, the dataset $z_i$ could contain this information, and in addition the geographical coordinates of its center point and of its vertices, its surface area, and potentially other exogenous covariate data. For instance, we could allocate each polygon with a category among the following three options: \emph{crop} ($x_i=1$), \emph{natural habitat} ($x_i=2$) or \emph{other} ($x_i=0$). 
For linear segment objects, the data $z_i$ could contain the geographical coordinates of its endpoints, and potentially other exogenous covariate data. A linear segment could be allocated with a category among \emph{single-species hedgerow} ($x_i=1$),   \emph{mixed-species hedgerow} ($x_i=2$) or \emph{no hedgerow} ($x_i=0$). 
In the case $\ell_i=1$ with only a single category $x_i=0$  no choice of allocation has to be made.  The space of all possible combinations of allocations is $\mathcal{X}=\mathcal{X}_1\otimes \mathcal{X}_2\otimes\ldots \otimes \mathcal{X}_n$. This finite collection contains $|\mathcal{X}|=\ell_1\times\ell_2\times\ldots\times \ell_n$ possible landscape allocations. If the geometric structure in $z_i$ can be supposed to be invariant through time, we describe temporal dynamics (if present) by the sequence $x_{i,t}$, $t=1,2,\ldots$ of categories over time. 

\subsection{Network model of interactions}\label{method_net}

We use a graphical representation of landscape to capture spatial or functional adjacency of landscape elements. Interaction between objects is modeled through a multilayer or multiplex network, i.e., a set of interacting single-network layers \citep{Boccaletti2014, Kivela2014}. Each layer corresponds to an object type, and each single-network layer represents the interaction among objects of same type. Interactions between different network layers represent interactions between objects of different type. 

\begin{figure}
\centering
\includegraphics[scale=0.4]{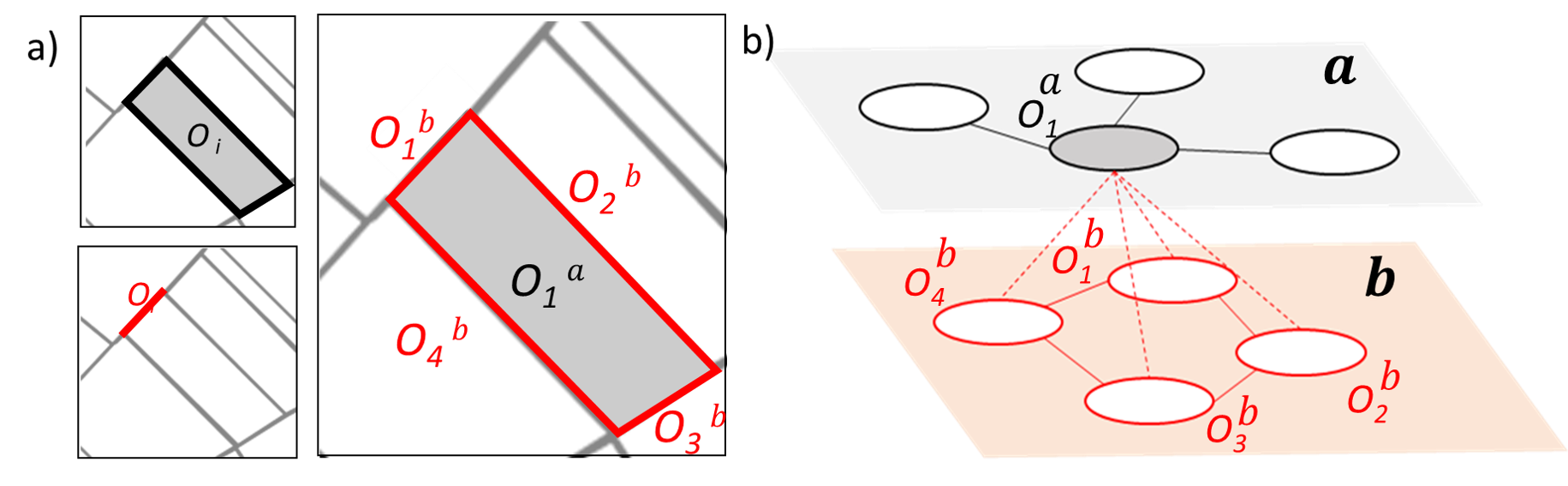}
\caption{Landscape representation. a) Polygon objects (patches, in grey) and linear segment objects (in red). b) Multi-layer network of interactions. Layer $a$: single network of interactions between patches; layer $b$: single network of interactions between linear elements; links between $a$ and $b$ represent interactions of patches and linear elements.}
\label{fig:Fig1}
\end{figure}

The focus of our models is on patches and linear landscape elements, such that we define a collection of objects with two types, $\textbf{o} =(\textbf{o}^a, \textbf{o}^b)$  (Figure  \ref{fig:Fig1}a), where  $o_i^a=(z_i^a,x_i^a)$ represents an object of patch type (layer $a$), and $o_i^b=(z_i^b,x_i^b)$ represents an object of linear segment type (layer $b$), see Figure  \ref{fig:Fig1}b.
The multilayer network is given as $ \mathcal{M} = ( \mathcal{G}, \mathcal{C})$, where $ \mathcal{G} = \{ G^a, G^b \} $ is the set of graphs defined over layers $a$ and $b$, respectively (Figure \ref{fig:Fig1}b). Layer $a$ is represented by the graph $G^a = (O^a, E^a)$ and describes the interaction among patches: nodes $O^a = (o_1^a, o_2^a, \ldots, o_{n^a}^a )$ represent $n^a$ patches, and links $E^a = (e_1^a, e_2^a,\ldots, e_{n_{E}^a}^a )$ represent the intra-layer interaction corresponding to the $n_{E}^a$ patch interactions. For the models in this paper, we assume that two patches interact if they are adjacent, that is, if they share part of their boundary. The  layer $b$ has similar structure and describes the interaction among linear landscape elements: nodes stand for linear landscape elements $O^b = (o_1^b, o_2^b,\ldots, o_{n^b}^b )$, and links stand for linear landscape element interaction $E^b = (e_1^b, e_2^b,\ldots, e_{n_{E}^b}^b )$. In this paper, we assume that two  linear elements interact if they intersect or have a vertex in common. 
The interaction between the single network layer $a$ and $b$ is described by the set of interconnections $ \mathcal{C} = \{ E_{ab}\} $ where $E_{ab} \subset O_a \times O_b$. In our framework, we assume a direct interaction among a node of $a$ and a node of $b$ when a linear element (i.e. $o_i^b$) is located on the boundary of a patch (i.e. $o_j^a$).

The full interaction structure is encoded in the adjacency matrix $\mathcal{A}$:
\begin{equation}
\mathcal{A} = \left(\begin{matrix}
A^a & A^{ab} \\
A^{ba} & A^b
\end{matrix}
\right), \quad \left(A^{ba}\right)^T=A^{ab},
\end{equation}
 where $A^a\in \mathbb{R}^{n^a \times n^a}$ and $A^b\in \mathbb{R}^{n^b \times n^b}$ represent the adjacency matrices of intra-layer interactions of  $a$ and $b$, respectively, and $A^{ab}\in \mathbb{R}^{n^a \times n^b}$ encodes inter-layer interactions among $a$ and $b$. For simplicity, we assume symmetry of interaction such that $\mathcal{A}=\mathcal{A}^T$, but the extension to asymmetric and directed interactions is straightforward. 
Entries of $\mathcal{A}$  could be binary to represent presence ($1$) or absence ($0$) of an edge, or may carry a weight value different from unity in case of presence of an edge. Non-binary weights could be based on distance between elements or on volumes/sizes of interacting elements.

Based on this landscape representation, we develop parametric models of probability distributions over the allocations $\bm x\in\mathcal{X}$, conditional on the information in $\bm z=(z_1,\ldots,z_n)$ and $\mathcal{A}$.
We  assume conditional independence of the value $x_i$ of object $o_i$ given the information from all adjacent elements: the landscape structure of non adjacent elements does not provide any information on object $o_i$ if we know all the adjacent objects. This framework allows for flexible interaction structures represented by the adjacency matrix $\mathcal{A}$ with sparse structure, such that we store only the positions and values of the relatively small number of non-zero entries. 
We formalize the conditional independence assumption through the following property of  equality in distribution of conditional probability distributions:
\begin{equation}\label{eq:condind}
    o_{i}\mid \mathcal{O}\setminus\{o_{i}\} \quad \stackrel{d}{=} \quad  o_{i}\mid \{o_j\in \mathcal{O}\mid (o_{i},o_{j})\in E \}
\end{equation}
where $E=E^a\cup E^b\cup E^{ab}$. We adopt notations such as $\bm o_{-i}$ to refer to the set of all objects $\mathcal{O}\setminus\{o_{i}\}$ with $o_i$ removed.

\subsection{Exponential family models for landscape descriptors}
\label{sec:expfam}

We set up a general probabilistic modeling framework based on Gibbs energies to define \emph{exponential family models}. 
For simplicity of notation, we will not distinguish between the full object information $o_i$ and the allocated value $x_i$, and we use $x_i$ henceforth.
For defining a model, we use $m$ functions $T_k:\mathcal{X}\rightarrow(-\infty,\infty)$, $k=1,\ldots,m$, measuring the value $T_k(\bm x)$ of summary statistics for  allocations $\bm x$. We refer to these functions as \emph{landscape descriptors}. We now define the probability of observing the allocation $\bm x$ as follows: 
\begin{equation}
p(\bm x) = \frac{1}{c(\bm\beta)}\exp\left(-\sum_{k=1}^m \beta_k T_k(\bm x)\right), \quad \bm x \in \mathcal{X}, \quad \bm\beta\in\mathbb{R}^m.\label{eq:efm}
\end{equation}
The normalizing constant $c(\bm\beta)>0$ is defined as
\begin{equation}\label{eq:normconst}
c(\bm\beta)=\sum_{\bm x \in \mathcal X}\exp\left(-\sum_{k=1}^m \beta_k T_k(\bm x)\right),
\end{equation}
such that probabilities in \eqref{eq:efm} sum up to $1$.
Since the number of possible configurations $|\mathcal{X}|$ is finite, the sum in Equation~\eqref{eq:normconst} is finite and the model is well-defined.
In practice, the number of possible configurations in $\mathcal{X}$ is usually very large, such that it is impossible to calculate the value of the  constant $c(\bm \beta)$.

Equation \eqref{eq:efm} defines a \emph{global specification} of the model where we consider the full allocation $\bm x$. In contrast, we also consider the local allocation probability of a landscape category $x_i$ conditional to the other landscape elements. Therefore, we  determine the probability of observing the category $x_i$ given the allocation of all the other elements $\bm x_{-i}=(x_1,x_2,\ldots,x_{i-1},x_{i+1},\ldots,x_{n})$, where we use the notation $(\bm x_{-i},y)=(x_1,\ldots,x_{i-1},y,x_{i+1},\ldots,x_{n})$. Then, the conditional probability is given as
\begin{equation}
p(x_i\mid \bm x_{-i}) = \frac{p(\bm x)}{\sum_{y\in \mathcal{X}_i} p(\bm x_{-i},y)}=\frac{\exp\left(-\sum_{k=1}^m \beta_k T_k(\bm x)\right)}{\sum_{y\in\mathcal{X}_i}\exp\left(-\sum_{k=1}^m \beta_k T_k(\bm x_{-i},y)\right)},
\label{eq:cond}
\end{equation}
and the unknown normalizing constant $c(\bm\beta)$  cancels out in expression~\eqref{eq:cond}.

\subsection{Examples of parametric models}
\label{sec:param}

Landscape descriptors are used to capture important landscape characteristics and features.  In \emph{composition terms}, such functions are the sum of contributions of individual objects; in \emph{spatial interaction terms}, we add up contributions that measure the interaction between two adjacent objects, and in \emph{time dependency terms} we compare configurations over two consecutive time steps.  An example specification is as follows, with $3$ spatial landscape descriptors given by
\begin{eqnarray}\label{eq:landesc}
T_1(\bm x)&=&\sum_{i\in O^a} t(x_i^a),\\
T_2(\bm x)&=&\sum_{( o_{i}^a,o_{j}^a)\in E^{aa}} t(x_i^a, x_j^a),\\
T_3(\bm x)&=&\sum_{( o_{i}^a,o_{j}^b)\in E^{ab}} t(x_i^a, x_j^b ),
\end{eqnarray}
and a descriptor for temporal dynamics:
\begin{equation}
    T_4(\bm x) \, = \,\sum_{i\in O^a} \sum_{\tau=2}^{H} t(x_{i,\tau}^a, x_{i,\tau-1}^a).
\end{equation}
Then, $T_1$ is a generic composition term, $T_2$ is a generic interaction term for objects of type $a$, $T_3$ is a generic interaction term for objects of different type, and $T_4$ describes time dynamics of objects of type $a$ over the time horizon $H$.
Table \ref{LDtable} illustrates concrete choices to specify the contributions to landscape descriptors involving two object types $a$ and $b$, here taken as patches and linear elements. For each object type, we allow for $2$ allocation  categories  $\mathcal{x}_i\in \{0,1\}$, such as \emph{crop} ($x_i^a=1$) or \emph{natural habitat} ($x_i^a=0$) and \emph{hedgerow} ($x_i^b=1$) or \emph{no hedgerow} ($x_i^b=0$) for patches and linear elements, respectively. 

\emph{Activity terms} are specific composition terms that provide direct control over the number of objects of a category by setting $T(\bm x)$ equal to the number of objects in specific category. To ensure identifiability, we fix a reference category (e.g., $x_i^a=0$ for objects of type $a$) and specify the activity term and its coefficient $\beta^a_{x_i^a}\in\mathbb{R}$ only for categories $x_i^a\not=0$, such that it is expressed relative to $x_i^a=0$. Implicitly, we have $\beta^a_{x_i^a=0}=0$. A positive coefficient $\beta^a_{x_i^a=1}>0$ would give relative preference to category $1$ over category $0$, such that landscapes tend to have more objects of category $1$ than of category $0$ for type $a$, provided that the energy terms of the other landscape descriptors do not influence the proportion of categories. 

\begin{center}
\begin{table}[h!]
      \caption{Examples of landscape descriptor terms.
      Notations: $\mathbb{Q}_p$ refers to the (empirical) $p$ percent quantile; $\mathbb{E}$ is the (empirical) expected value. }

\renewcommand{\arraystretch}{1.5}    
\begin{tabular}{| >{\centering\arraybackslash}m{3cm}| >{\centering\arraybackslash}m{2cm}|>{\centering\arraybackslash}m{10cm}| @{}m{0pt}@{}}
\hline
    \multirow{6}{*}{Composition} & \centering Activity term & 
    $
 t(x_i^a)=
 \left\{
 \begin{array}{l l}
    1  & \text{ if } x_i^a = 1 \\
    0  & \text{ otherwise}
 \end{array}
 \right. 
$  \\

    \cline{2-3}
     & \centering  Patch area & 
     $t(x_i^a; p)=
 \left\{
 \begin{array}{l l}
    1  & \text{ if } x_i^a = 1, \mathrm{area}(o_i^a) \leq \mathbb{Q}_{p}(\mathrm{area}(o_i^a)) \\
    0  & \text{ otherwise}
 \end{array}
 \right.
$\\

     \cline{2-3}
     & \centering Long segments  & 
     $   t(x_i^a)=
 \left\{
     \begin{array}{l l}
    1  & \text{ if } x_i^b = 1,\ \mathrm{length}(o_i^b) \geq \mathbb{E}[\mathrm{length}(o_i^b)] \\
    0  & \text{ otherwise}
    \end{array}
    \right.
    $\\
 \cline{2-3}
     & \centering Horizontal segments   & 
     $   t(x_i^a)=
 \left\{
     \begin{array}{l l}
    1  & \text{ if } x_i^b = 1, \mathrm{angle}(o_i^b)\in [0,\frac{\pi}{6}]\cup \left[\frac{5\pi}{6},2\pi\right]
    
    \\
    0  & \text{ otherwise}
    \end{array}
    \right.
    $\\
     \hline
   
    \multirow{3}{*}{\shortstack[l]{Interaction \\ (Adjacency)}} & \centering Patch-patch  & 
    $t(x_i^a, x_j^a )=a^{a}_{ij} $ \\
    
    \cline{2-3}
     & \centering Segment-segment & 
     $ t(x_i^b, x_j^b )=a^{b}_{ij} $ \\
    
    \cline{2-3}
     & \centering Patch-segment & 
     $    t(x_i^a, x_j^b )=a^{ab}_{ij}$ \\

     \hline
     
    \multirow{1}{*}{Time dependency} & \centering Patch rotation & 
    $    t(x_{i,\tau}^a, x_{i,\tau-1}^a)=
 \left\{
     \begin{array}{l l}
    1  & \text{ if } x_{i, \tau-1}^a = x_{i, \tau}^a  \\
    0  & \text{ if } x_{i, \tau-1}^a \neq x_{i, \tau}^a 
     \end{array}
    \right.
    $  \\
    
     \hline
     
\end{tabular}
\label{LDtable}
\end{table}
\end{center}

\subsection{Iterative landscape simulation algorithm}\label{method_sim}

We implement a Metropolis-Hastings (MH) algorithm to iteratively simulate a discrete-time Markov chain whose stationary distribution corresponds to the target model \citep[e.g.,][]{Descombes2013} where the configuration of the allocated land-use categories $\bm x$ is the state variable of the system. The four  main steps of MH-MCMC are as follows: i) define an initial state $\bm x^{(0)}$; and then iteratively ii)  propose a new state $\tilde{\bm x}$ given the current state $\bm x^{(j)}$, based on a proposal distribution $q(\tilde{\bm x}\mid \bm x^{(j)})$, and iii) decide on acceptance ($\bm x^{(j+1)}=\tilde{\bm x}$) vs. rejection ($\bm x^{(j+1)}=\bm x^{(j)}$)  based on checking $U_j<R_j$ where $U\sim\mathrm{Unif}(0,1)$,  and the acceptance ratio $R_j$ is given by
\begin{equation}\label{eq:hr}
  R_j=\frac{p(\tilde{\bm x}) q(\bm x^{(j)}\mid \bm \tilde{\bm x}) }{p(\bm x^{(j)}) q(\tilde{\bm x}\mid \bm x^{(j)})};
\end{equation}
finally,
iv) Return the final configuration $\bm x^{(N_0)}$ after $N_0$ iterations. A more detailed schematic overview is given in the Supplementary Material.
If we need more than one realization of the landscape, we can either run chains in parallel, or we may run a single chain but return a sample given by the  states indexed by $N_0+\ell N$, $\ell=1,2,\ldots$, with the burn-in period $N_0>0$ and $N-1$ intermediate states left out to avoid autocorrelation in the final sample. 
Since the parameter vector $\bm\beta$ of the model is fixed during each MCMC run, the intractable normalizing constant $c(\bm \beta)$ in \eqref{eq:efm} cancels out in the acceptance ratio. However, we  have to update the calculation of the set of landscape descriptors for each new configuration $\tilde{\bm x}$ during the iterations.  

With respect to the choice of the initial state $\bm x^{(0)}$ of the system, we have to ensure that $p(\bm x^{(0)})>0$, and that valid paths $\bm x^{(1)},\ldots,\bm x^{(j)}$ to any configuration $\bm x^{(j)}$ with $p(\bm x^{(j)})>0$ are proposed with strictly positive probability. The models presented in Section~\ref{sec:param} and implemented in the remainder of the paper satisfy $p(\bm x)>0$ for all possible configurations $\bm x$, such that any initial state is valid. In more general cases, hereditary properties when moving between configurations must be checked. We initialize the system  state either at random (draw the category of an object $o_i$ among all $\ell_i$ available categories), or by attributing only one category for each object type,  or by using an observed configuration from real data. 
For the proposal  of a new state, we use the following default proposal distribution $q(\cdot\mid \cdot)$ of random walk type. We first select an object at random. In case of temporal dynamics, the time step is also selected at random. Then, we propose a new category for this object, different from the current one and drawn at random. This proposal distribution leads to symmetry of proposals to move forward and backward, such that the ratio $q(\bm x^{(j)}\mid \bm \tilde{\bm x})/q(\tilde{\bm x}\mid \bm x^{(j)})$ cancels out from \eqref{eq:hr}. 

The landscape descriptors $T_k$ are sufficient statistics in an exponential family model --  they contain all the information on $\bm \beta$ that we can draw from $\bm x$. Therefore, we can monitor the convergence of Markov chains to their stationary distribution by  checking the $m$ simulated series $T_k^{(j)}$ through trace plots, which further allows us to determine the burn-in period, and to analyze the mixing behavior of the Markov chains to fix the number $N-1$ of intermediate states to be left out \citep[see, e.g.][]{Kieu2013}. In practice, we have found that the number of iterations needed for burn-in strongly depends on the combination of size of the landscape and  complexity of the model, and especially on the type of landscape descriptors involved. The running time necessary to simulate one landscape in one Markov chain for the examples discussed in this paper ranges from several seconds to several minutes.

\subsection{Simulation examples}
We illustrate the simulation algorithm for the subdomain D1 presented in Section~\ref{sec:data}, and we visually explore how landscape simulation output changes when varying parameters $\beta_k$ in \eqref{eq:efm}. We focus on object interactions: \textit{crop-hedge adjacency} ($\beta_4^a$), \textit{hedge-hedge adjacency} ($\beta_3^b$), and \textit{crop-hedge adjacency} ($\beta_3^a$)), as defined in Table~\ref{LDtable}. All other descriptors have coefficient $\beta_k=0$, i.e., are not present in the model.

We simulate three  landscape models, each one having one of the three $\beta$-coefficients different from $0$, with values $-1, -0.33, 0.33, 1$.  Each simulation is run in a separate Markov chain and takes about 10 seconds for the \emph{small regions D1} and \emph{D2} and it takes about 30 minutes for the \emph{large region D3}. Figure 3 in the supplementary material shows some traceplots of the landscape descriptors and confirms fast mixing of the chains. We have finally fixed relatively large numbers of burn-in steps of $N_0=10,000$ for the small domains D1 and D2, and of $N_0=1\times 10^6$  for the large domain D3, to ensure that the chains have reached the stationary distribution. 

\begin{figure}
\centering
\includegraphics[width=\textwidth]{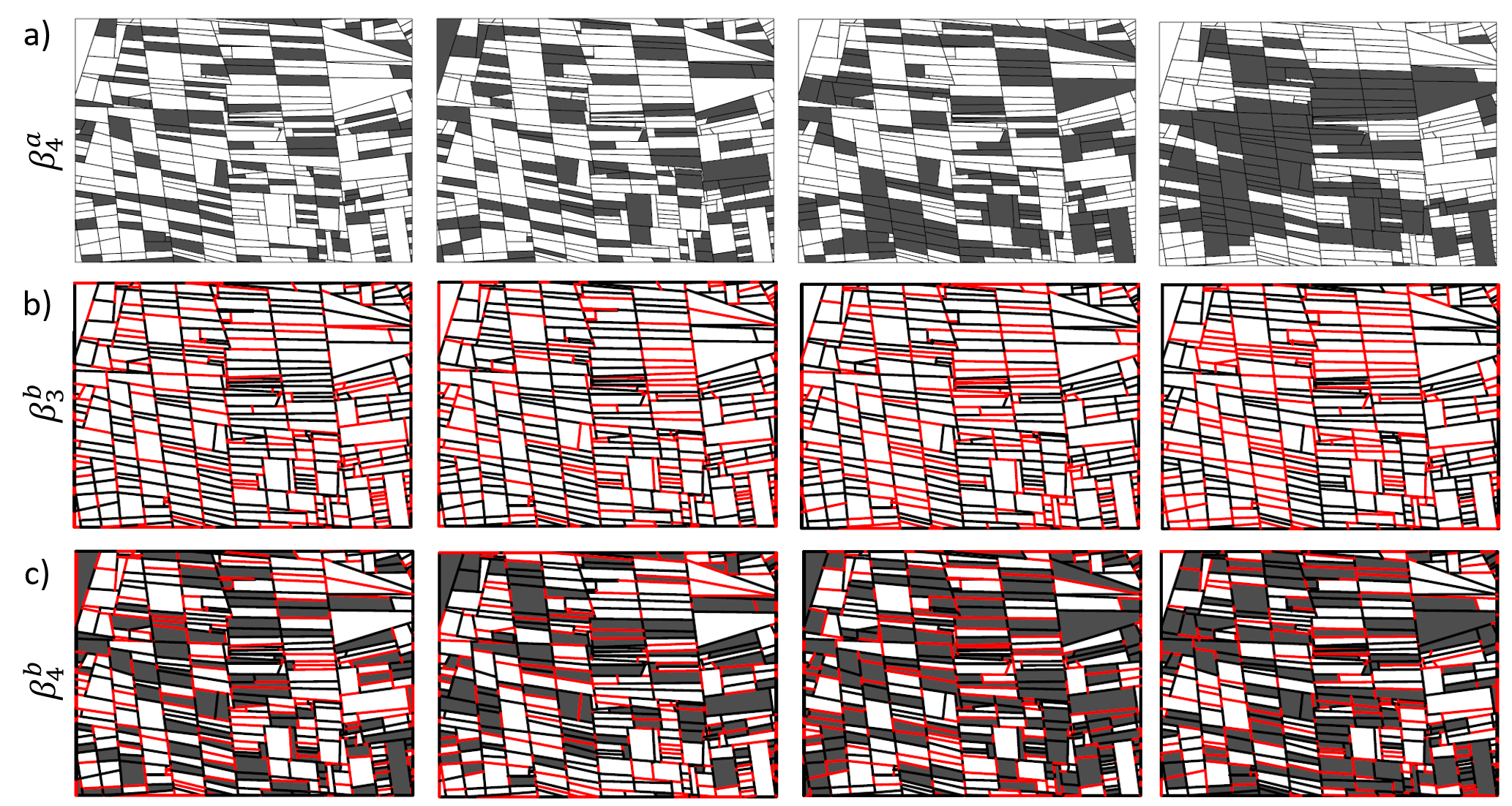}
\caption{Landscape simulations on D1. Panel a):  varying \textit{crop-crop adjacency}; Panel b): varying \textit{hedge-hedge adjacency}; Panel c): varying \textit{crop-hedge adjacency}. Columns from left to right: coefficient $-1, -0.33, 0.33, 1$. }

\label{fig:Sim_Ex}
\end{figure}

Figure \ref{fig:Sim_Ex} shows one simulation for each configuration of interaction type and coefficient value.  A negative coefficient for the interaction among elements of the same type (Figure \ref{fig:Sim_Ex}a and \ref{fig:Sim_Ex}b) yields a fragmented allocation of crops (Figure \ref{fig:Sim_Ex}a) and hedges (Figure \ref{fig:Sim_Ex}b), respectively, while a positive coefficient results in clustered configurations. As for the multiplex interaction of \textit{crop-hedge adjacency} in Figure \ref{fig:Sim_Ex}c, a negative coefficient leads to  hedges being located away from  crop-patch boundaries, while they concentrate on such boundaries for a positive parameter.

\section{Statistical inference and model validation}\label{method_Stat_inf}
\subsection{Parameter inference}
\label{sec:inference}
To infer the allocation mechanism of a real landscape, we estimate  the parameter vector $\bm\beta$.  The likelihood function is not tractable in practice due to the normalizing constant $c(\bm\beta)$ in the probability mass function \eqref{eq:efm}. Instead, we construct a pseudo-likelihood based on conditional distributions (see \citet{Besag1974,Baddeley2000}). 
 Given $n$ objects  $\bm x=(x_1,\ldots,x_n)$  with their allocation categories, we define the pseudo-likelihood as the product of the conditional probability of the category  $x_i$  given all the other variables $\bm x_{-i}$; i.e., it is the composite likelihood of conditional distributions \citep{Besag1974, Varin2011} given as
\begin{equation}\label{eq:LL}
\mathcal{L}=  \prod_{i=1}^{n}p(x_{i}\mid \bm x_{-i})
\end{equation} 
where the conditional probability $ p(x_{i}\mid \bm x_{-i})$ is defined in Equation \eqref{eq:cond}. Therefore, the unknown normalizing constant $c(\bm\beta)$  cancels out.
We assume that direct interaction of landscape objects occurs only along the edges of the multiplex-network graph, such that we only need the information from adjacent objects in $\bm x_{-i}$ owing to conditional independence. 

If $x_i$ has only two possible allocation categories $\mathcal{X}_i=\{0,1\}$, then we can write $\tilde{\bm x}$ for $\bm x$ with $x_i$ replaced by the alternative level, and \eqref{eq:cond} is equivalent to the logistic regression equation
\begin{equation}\label{eq:logreg}
\log \frac{p(x_{i}\mid \bm x_{-i})}{1-p(x_{i}\mid \bm x_{-i})} = \sum_{k=1}^m \beta_k \left(T_k(\bm x)-T_k(\tilde{\bm x})\right).
\end{equation}
We exploit the closed-form conditional probabilities \eqref{eq:cond} and \eqref{eq:logreg} for parameter estimation of $\bm \beta$ using standard software implementations of logistic regression (if $\ell_i=2$), or using the more general pseudo-likelihood framework (if $\ell_i>2$).

We implement parametric bootstrapping for statistical inference and model selection.  The maximum pseudo-likelihood estimator $\widehat{\bm\beta}$ is known to be asymptotically consistent and normal \citep{Jensen1991,Varin2011}, but inference is more involved. Specifically, asymptotic variance-covariance parameters are more difficult to obtain from the model fit as compared to  full likelihood, and estimation bias is possible with finite-sample data. Instead, we use our simulation algorithm for parametric bootstrapping to validate good estimation performance of the pseudo-likelihood, specifically  unbiasedness, and for obtaining  parameter confidence intervals.   

In practice, we generate $n_{\text{boot}}$ landscapes from the  fitted model using $\bm\beta = \widehat{\bm\beta}$ to obtain a sample $ \widehat{\bm\beta}_1^*, \dots  \widehat{\bm\beta}_{n_{\text{boot}}}^* $ from the pseudo-likelihood estimator, and then check for estimation bias and derive Monte-Carlo confidence intervals. For a test of the null hypothesis $\beta_k=0$ for fixed $k\in\{1,\ldots,m\}$, i.e. to check if the landscape descriptor $T_k$ is significant, we implement a Monte-Carlo test where we sample from the fitted model, but with the modification $\hat{\hat{\beta}}_k=0$. The null is rejected if the value $\hat{\beta}_k$ does not lie within the one-sided Monte-Carlo confidence interval of $\hat{\hat{\beta}}_k$, i.e., if less than $\alpha\%$ (e.g., $\alpha=5$) of the $\beta_k$-values estimated for the simulations have the same sign and higher absolute value than the value estimated for the data \citep[see, e.g.,][]{Davison1997}. 

\subsection{Validation metrics based on landscape structure}\label{sec:valid}
For statistical validation of models, it is important to check if the distribution of landscape descriptors -- as produced by the fitted model -- is in line with the observed value of this descriptor. We develop such checks in the data application in Section~\ref{method_BVD}. 

Moreover, various metrics exist to assess if simulated landscape patterns succeed in measuring the landscape functionality and different ecological relevancy in diverse applications \citep{Kupfer2012, Frazier2017}. Such metrics are often strongly correlated. It is important to assess if a model such as the one we propose in  \eqref{eq:efm}, usually characterized through a small number $m$ of landscape descriptors,  is capable to generate metric values similar to those in the observed data.  Some of commonly used landscape metrics focus on a  patch-mosaic model as in our work (i.e., the landscape is simplified into a mosaic of discrete habitat patches). Many metrics have been developed for landscapes conceptualized as an environmental gradient  \citep[i.e., for raster representations, see][]{Mcgarigal1995, Cushman2010}.
Here, we assess how diverse spatial patterns in data are reproduced by our model through metrics based on graph theory (\emph{network metrics}), often applied to patch-mosaic model \citep{Urban2001, Minor2008, Urban2009, Lu2016}, or through metrics based on gradient theory \emph{raster metrics} \citep{Mcgarigal1995}, where we transform the patch-mosaic output in our simulation into a raster. We compare the metrics calculated for the real landscape to the empirical distribution of metrics from the simulated landscapes, the latter obtained through the parametric bootstrap. The metrics are presented in Table \ref{tab:metrics}. We focus on standard network metrics, which are intuitive and useful in different application contexts \citep{Urban2001, Minor2008}. 
Network metrics are evaluated either at the node scale (with one value per node) or at network scale. Node scale helps to identify vital nodes associated with structural or functional objectives \citep{Lu2016}, while network scale summarizes the whole topology \citep{Urban2001, Calabrese2004}. From gradient theory, we compute metrics reviewed by \cite{Mcgarigal1995}, and we follow \cite{Cushman2008} by choosing the metrics identified as ``highly universal and consistent class-level landscape structure components".

For the raster analysis, we use the \texttt{R} package \texttt{raster} \citep{Hijmans2015} to transform the spatial objects (i.e., polygons and linear segments) into pixels with coordinates \textit{x}, \textit{y} and a categorical value and the package \texttt{landscapemetrics} \citep{landscapemetrics} to evaluate the raster metrics. In our example with two polygon types (crops and hedges) and two edge types (hedge or not), we obtain 3 categories (also called habitats):  crop, semi-natural habitats, and hedge. Here, absence of hedges is not a class in itself but rather leads to a lack of barriers. 

\begin{table}[h]
\centering
\caption{Landscape metrics. A star $^\star$ indicates metrics that we normalize when comparing networks with different node numbers. Support is either ``node" for node-scale network metrics, ``network" for global network metrics, or ``raster" for grid-based metrics. References: [1] \cite{Urban2009}, [2] \cite{Lu2016}, [3] \cite{ Latora2001}, [4] \cite{Mcgarigal1995}, [5] \cite{Cushman2008}  }

\label{tab:metrics}
\resizebox{\textwidth}{!}{%
\begin{tabular}{|c|>{\centering\arraybackslash}p{8cm}ccc|}
\hline
\textbf{Name} & \textbf{Description} & \textbf{Support} & \textbf{Range} & \textbf{Reference} \\ \hline
Degree$^\star$ & Number of directly connected node neighbors & node & {[}0, 1{]} & [1],[2] \\
Coreness & K-shell decomposition for node’s spreading influence & node & {[}0,$\infty$) & [1],[2] \\
Degree grade 2$^\star$ & Number of nodes at most $2$ away from given node 
& node & {[}0, 1{]} & [1],[2] \\
Eccentricity$^\star$ & Maximum shortest path to nodes & node & {[}0, 1{]} & [1],[2] \\
Closeness & Reciprocal of total length of shortest paths to other nodes\ & node & {[}0,$\infty$) & [1],[2]\\
Betweenness$^\star$ & Potential power to control information flow & node & {[}0, 1{]} & [1],[2] \\
Diameter & Longest path & network & {[}0,$\infty$) & [1] \\
Efficiency & Efficiency of information exchange & network & {[}0,$\infty$) & [2],[3] \\
Cluster avg & Proportion of interconnected adjacent nodes of a vertex & network & {[}0,1{]} & [2],[3] \\
PLAND {[}\%{]} & Percentages of habitats in the landscape & raster & {[}0,100{]} & [4],[5] \\
PD {[}\# / ha $\times$ 100{]} & Patch density & raster & {[}0,$\infty$) & [4],[5] \\
ENN {[}m{]} & Mean nearest neighbor distance & raster & {[}0,$\infty$) & [4],[5] \\
PARA {[}/{]} & Perimeter-area ratio & raster & {[}0,$\infty$) & [4],[5] \\
IJI {[}\%{]} & Interspersion/juxtaposition index measuring spatial  intermixing of different habitats & raster & {[}0,100{]} & [4],[5] \\
CLUMPY {[}/{]} & Clumpiness index measuring deviation from randomness & raster & {[}-1,1{]} & [4],[5] \\ \hline
\end{tabular} %
}
\end{table}

\section{Application to the Lower Durance Valley in southern France}\label{method_BVD}
\subsection{Landscape model structure}
We fit parametric stochastic models in the study area, analyze them to select the most appropriate descriptors, and run a simulation experiment to study how well landscape metrics not directly controlled through the Gibbs energy terms are reproduced.  
We define a full model based on a moderate number of landscape descriptors we deem potentially useful to appropriately characterize the land-use allocation mechanism, and we estimate its parameters using the composite likelihood  of Section~\ref{sec:inference}. We then run parametric bootstrap to obtain confidence intervals and test the relevance of individual landscape descriptor. For patches and linear elements, we consider $2$ categories in each case: \emph{crop}, \emph{semi-natural area} for patches; presence or absence of a \emph{hedgerow} allocated on a patch border for linear elements.
We apply the full model described in Equation~\eqref{eq:efm} using the following landscape descriptors.
For patch objects: \textit{activity parameter} ($T_1^a$), \textit{patch area} ($T_2^a$) using a ${25\%}$-quantile,  \textit{crop-hedge adjacency} ($T_3^a$), \textit{crop-crop adjacency} ($T_4^a$), with coefficients $\beta_2^a$, $\beta_3^a$, $\beta_4^a$, respectively; for linear elements: \textit{activity parameter} ($T_1^b$), \textit{long segments}  ($T_2^b$), \textit{hedge-hedge adjacency} ($T_3^b$), \textit{allocation of horizontal segments} ($T_4^b$), with coefficients $\beta_2^b$, $\beta_3^b$, $\beta_4^b$, respectively, with formulas given in Table \ref{LDtable}. 
We select a condition on the average length of hedges to test if there is  allocation preference towards hedges shorter or longer than the average. For areas, we instead use a condition on the first quartile since the patch area distribution shows high variance 
and small field sizes may benefit biodiversity through easier access to adjacent fields with complementary resources \citep{Sirami2019}. In our landscape, crops tend to be allocated on moderately sized patches while big patches are dedicated to open area, and so we focus on small patch allocation. However, we can extend the model with another descriptor to control large patch allocation through a condition using the $75\%$-quantile of patch size ($T_{2, q75}^a$)) to compare the performance of the simple model ($D1$) and the extended model ($D1+$) for domain D1. We point out some salient results of the comparison of $D1$ and $D1+$, while detailed numerical results  are reported in the Supplemental Material. Time dynamics such as crop rotation cannot be estimated here due to lack of data.

\subsection{Parameter inference}
We use the logistic regression \eqref{eq:logreg} proposed in Section~\ref{method_Stat_inf} to estimate the category allocation mechanism of patches and segments for the three regions D1, D2, D3.  Table~\ref{Estime_tab} reports parameter estimates of the coefficients $\bm\beta$ where standard errors, confidence intervals, significance (with respect to the null $\beta_k=0$) based on a parametric bootstrap with $100$ simulations. Figure 8 in the Supplemental Material reports the boxplots for the parameter estimates. 
Throughout, the estimators are unbiased. Estimated parameters are all significant with one exception for D1 concerning hedge attribution to relatively long segments and hedge and crop interaction for D1 and D2. The signs of all the estimates are the same across the three subdomains, implying the same trends, such that the domains are structurally similar.  \textit{Patch area} allocation to small patches has  negative coefficient, i.e. crop is usually not cultivated on very small areas. By contrast, we find a positive coefficient for \textit{ crop-crop adjacency}, meaning that crops are preferably located in spatial clusters. \textit{Crop-hedge adjacency} is not significant for the small domains D1 and D2 but for D3, and estimated coefficients are negative,  meaning that hedges tend not to be directly allocated around crop patches. \textit{Hedge-hedge adjacency} and the allocation of hedges to \textit{horizontal segments} have significant positive coefficients throughout, such that allocation of hedges tends to be clustered (i.e. continuous in space) with horizontal orientation for the purpose of breaking winds.  
In the extended  model $D1+$, the  coefficient controlling crop allocation to large patches is significantly negative, and results discussed below show major improvements in the model's capability to realistically reproduce patch-related landscape metrics leading to a satisfactory performance of the extended model; see the Supplemental Material for detailed results. Overall, estimated parameters have comparable magnitudes between the three domains. In D2, we discern a strong signal indicating many short, strongly horizontally oriented hedges, as compared to D1 and the larger domain D3.

\begin{table}
\centering
\caption{Parameter estimates for the subdomains  (first column) of the study area. C stands for ``crop" and H for ``hedge". ``Mean" and ``SD" values were calculated through a parametric bootstrap with $100$ simulations from each of the fitted models. Bold face indicates significance of the descriptor at the $95\%$ level using Monte-Carlo simulation for the test statistics under the null hypothesis. As for the model D1+, we estimate also the parameter related to Big Area: \textbf{-1.13}, -1.11, 0.29 for Estimated, Mean and SD, respectively. }
\label{Estime_tab}
\resizebox{\textwidth}{!}{%
\begin{tabular}{|rr|rrrr|rrrr|}
\hline
                                          &                   & \multicolumn{4}{c|}{Crop}                                                    & \multicolumn{4}{c|}{Hedge}                                                       \\ \cline{3-10} 
                                          &                   & Activity  & Small area     & C-H & C-C & Activity  & Long H    & H-H & Horizontal H \\ \hline
\multicolumn{1}{|c|}{\multirow{4}{*}{D1}} & Estimate         & \textbf{-1.08} & \textbf{-1.24} & -0.05                & \textbf{0.37}       & \textbf{-2.38} & -0.10          & \textbf{0.77}         & \textbf{1.58}          \\
\multicolumn{1}{|c|}{}                    & Mean              & -1.09          & -1.25          & -0.05                & 0.37                & -2.37          & -0.09          & 0.77                  & 1.58                   \\
\multicolumn{1}{|c|}{}                    & SD                & 0.37           & 0.34           & 0.09                 & 0.10                & 0.19           & 0.19           & 0.07                  & 0.19                   \\\hline
\multicolumn{1}{|l|}{\multirow{4}{*}{D1+}} & Estimate         & \textbf{-1.01} & \textbf{-1.52} & 0.03       & \textbf{0.40}       & \textbf{-2.38} & -0.10          & \textbf{0.77}         & \textbf{1.58}         \\
\multicolumn{1}{|l|}{}                    & Mean              & -1.06          & -1.55          & 0.05                & 0.39                & -2.37          & -0.09          & 0.77                  & 1.58                   \\
\multicolumn{1}{|l|}{}                    & SD                & 0.34          & 0.35          & 0.07                 & 0.10               & 0.19           & 0.19           & 0.07                  & 0.19                    \\ \hline
\multicolumn{1}{|l|}{\multirow{4}{*}{D2}} & Estimate         & \textbf{-1.65} & \textbf{-0.96} & -0.08                & \textbf{0.64}       & \textbf{-3.38} & \textbf{-0.58} & \textbf{0.77}         & \textbf{3.65}          \\
\multicolumn{1}{|l|}{}                    & Mean              & -1.67          & -0.95          & -0.08                & 0.64                & -3.40          & -0.58          & 0.77                  & 3.69                   \\
\multicolumn{1}{|l|}{}                    & SD                & 0.33           & 0.28           & 0.08                 & 0.08                & 0.25           & 0.19           & 0.09                  & 0.22                   \\ \hline
\multicolumn{1}{|l|}{\multirow{4}{*}{D3}} & Estimate         & \textbf{-1.87} & \textbf{-1.37} & \textbf{-0.15}       & \textbf{0.66}       & \textbf{-3.01} & \textbf{-0.16} & \textbf{0.95}         & \textbf{1.97}          \\
\multicolumn{1}{|l|}{}                    & Mean              & -1.88          & -1.39          & -0.15                & 0.65                & -3.02          & -0.17          & 0.95                  & 1.98                   \\
\multicolumn{1}{|l|}{}                    & SD                & 0.10           & 0.14           & 0.02                 & 0.03                & 0.06           & 0.05           & 0.02                  & 0.06                   \\ \hline

\end{tabular}%
}
\end{table}

\subsection{Comparison of observed and simulated landscape metrics}

We discuss graph- and raster-based metrics for crop allocation in the \textit{small region D1} by comparing observed values with the simulated distribution given the fitted model, using $100$ MCMC samples.  
This further allows testing the null hypothesis that the observed metric could have been generated by the model by using two-sided $95\%$  Monte-Carlo confidence  intervals. Results for hedges and for other domains (D2, D3) are structurally similar and can be found in the Supplemental Material.

\subsubsection{Network-related metrics}
\begin{figure}
\includegraphics[scale=0.35]{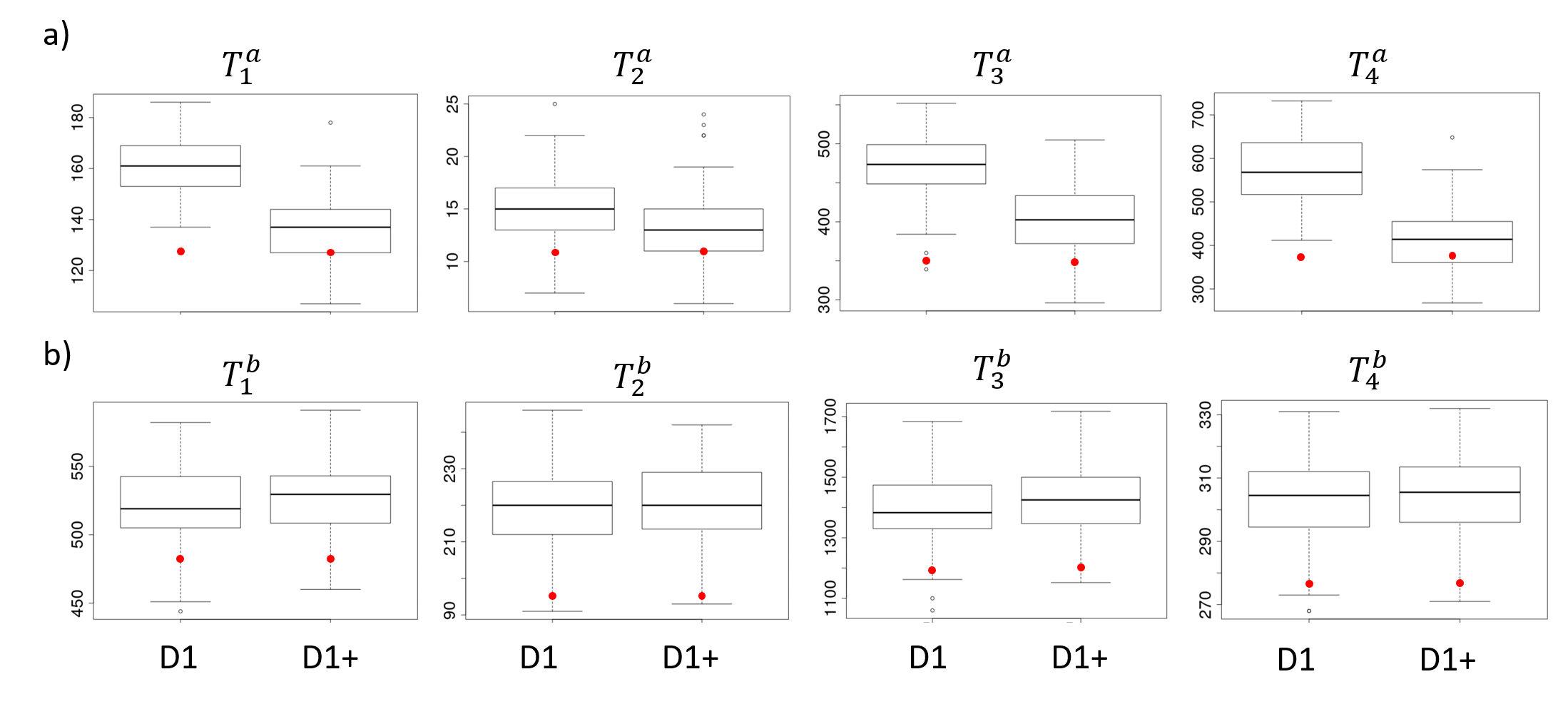}
\caption{Landscape descriptors for the \textit{small region D1} for the basic model (D1) and the extended model (D1+). Panel a) crop network; Panel b) hedgerow network. Boxplots represent simulated landscapes. The red dot represents the real landscape value.}
\label{fig:LandDscr}
\end{figure}

For models D1 and D1+, Figure \ref{fig:LandDscr} shows real and simulated  landscape descriptor values through the red dot and the boxplot, respectively. The models tend to appropriately reproduce the real landscape descriptor values, especially in D1+ where the control over large patch allocation leads improves behavior for all patch-related descriptors. 

\begin{figure}
\centering
\includegraphics[width=\linewidth]{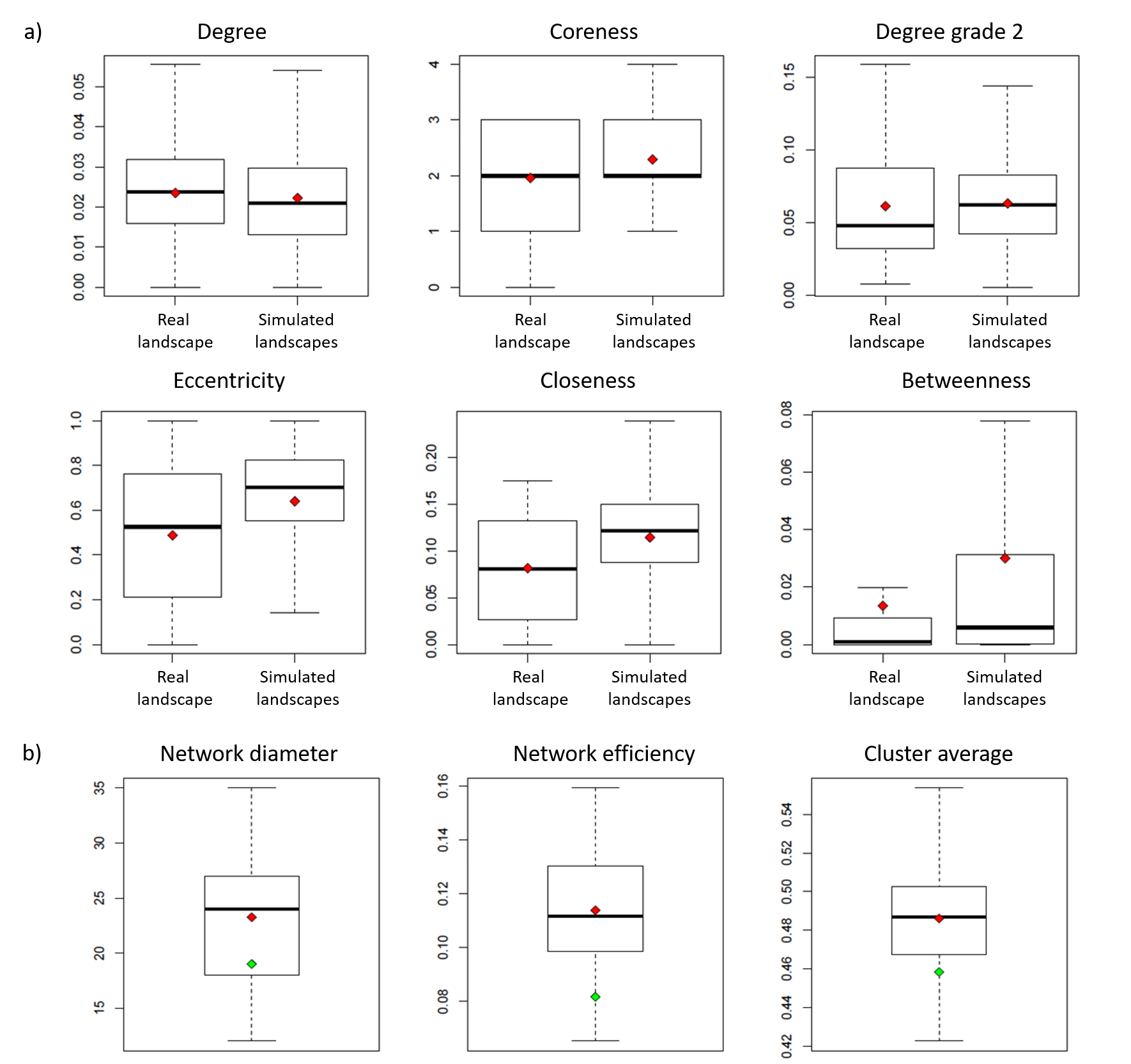}
\caption{Validation metrics for crop network. Panel a) metrics at node scale. Red dots represent the mean values of the node metrics of the real and simulated networks, respectively.  Panel b) metrics at network scale. Boxplots represent simulations, red dots represent mean values of simulations, green dots represent the observed value.}
\label{fig:Fig6}
\end{figure}

Figure \ref{fig:Fig6} shows  real and simulated network metrics for patches. 
For node-scale metrics (two top rows), we observe good overlap of boxplots for observed and simulated values, with the exception of \emph{Betweenness} where we tend to simulate larger values. \emph{Betweenness} evaluates the number of the shortest paths going through a node when connecting any two other nodes and is very heavy-tailed since it shows high variability among different networks. 
Differences in the boxplots (real vs. simulated) may in part be due to the $100$ times larger sample size in the case of simulations, while strong right-skewness and heavy tails are present in both cases. This skewness highlights that few crop patches have the bridging role of connecting different crop clusters, which is fundamental for maintaining the connectivity of the landscape \citep{Estrada2008, Urban2009, Belgrano2015}, and this property is preserved in the fitted model.

For network-scale metrics (last row of Figure~\ref{fig:Fig6}) we show the real landscape value within the boxplot of simulated values. The observed \textit{Diameter} is within the interquartile range of the simulated ones, while observed values of \textit{Efficiency} and \textit{Cluster Average} are located towards the lower extremes of simulations. The numbers in Table \ref{tab:MC_test} report the proportion $p\in[0,0.5]$ of the simulated metric values that are ``more excentric" than the observed one; e.g., if the observed value is below the median and $26$ among the $100$ simulated values are even lower, we report $0.26$.   These \emph{pseudo-$p$-values} imply that observed metrics for the crop network still appear realistic under the model. Overall, network-scale results indicate slightly stronger clustering of crop in the model as compared to reality, but still with similar order of magnitude for metric values. We also report pseudo-$p$-values for hedge network-scale metrics in Table \ref{tab:MC_test}, which show stronger discrepancy between observed and simulated values. Node-scale metrics for hedges, more directly controlled through the network descriptors included in our model, remain satisfying.
For the sake of parsimony and simplicity, the model studied here does not directly control descriptors evaluating the number or dimension of clusters, but only local relationships among patches. It is not surprising to obtain better performance of \textit{neighborhood-based centralities} in comparison to \textit{path-based centralities} and  metrics, the latter based on the whole landscape. 

\subsubsection{Raster metrics}
\begin{figure}
\centering
\includegraphics[width=\linewidth]{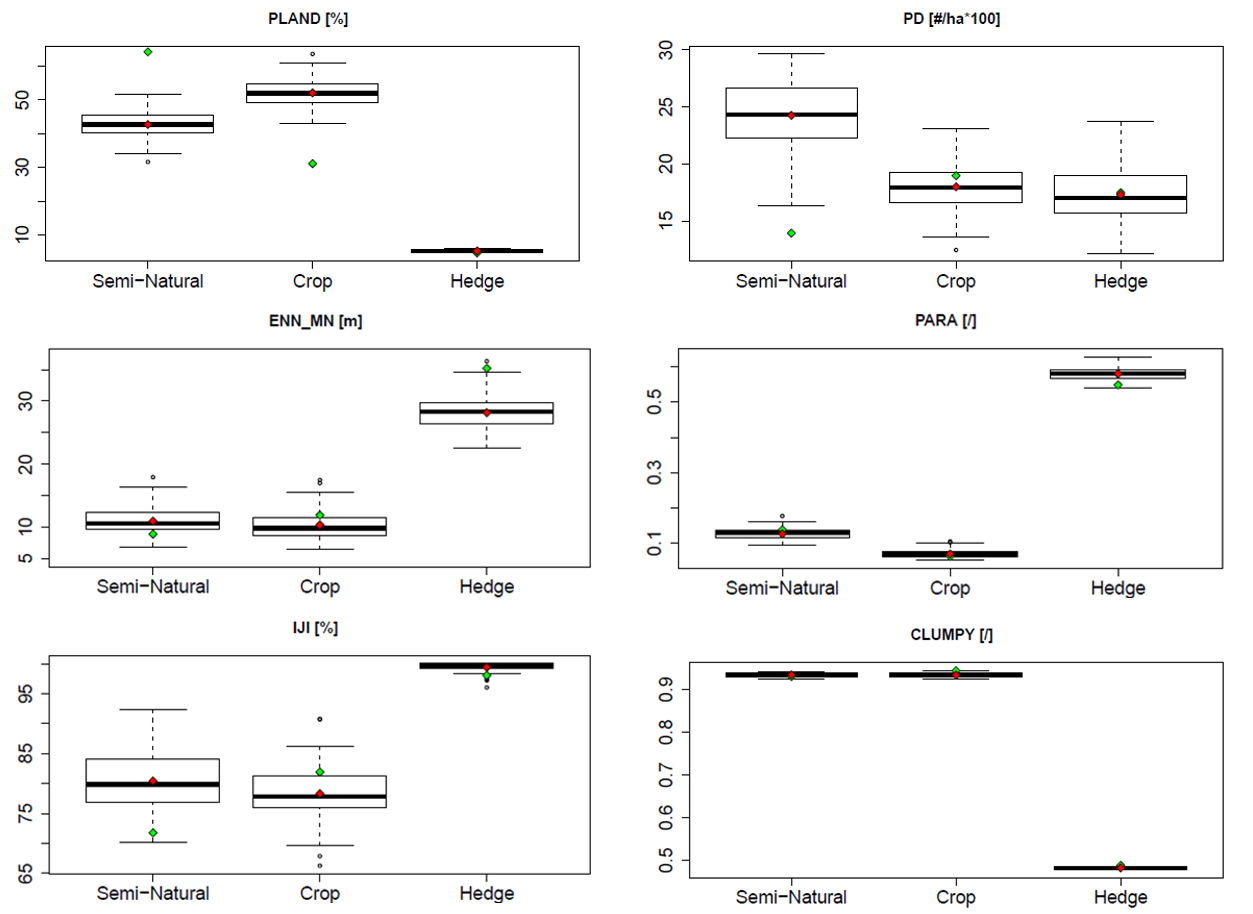} 
\caption{Raster-based landscape metrics for model D1. Boxplots: simulated values. Red dots: mean of simulated values. Green dots: observed value.}
\label{fig:Fig_lsm}
\end{figure}

Figure \ref{fig:Fig_lsm} shows the raster-based landscape metrics of FRAGSTAT; see Table~\ref{tab:metrics} for their description. The model has some difficulty to capture certain observed properties in this grid representation. In particular, the model directly controls only the number of elements according to certain criteria, but not proportions of spatial area. The pseudo-$p$-values  listed in Table~\ref{tab:MC_test} numerically confirm this mismatch between model and real landscape for the proportion of landscape categories (PLAND), and for Patch Density (PD) of semi-natural patches.
Other very small pseudo-$p$-values also indicate a mismatch, but  the boxplots in Figure~\ref{fig:Fig_lsm} show that the order of magnitude of metric values is still well captured by the model, even though the variability of simulations is not large enough. Moreover, in the raster representation two adjacent objects of the same type and category are considered as a single habitat, which is  different from the vector-based representation in the model.

\subsubsection{Correlation analysis of landscape descriptors and metrics}
\begin{figure}[h!] 
\includegraphics[width=\linewidth]{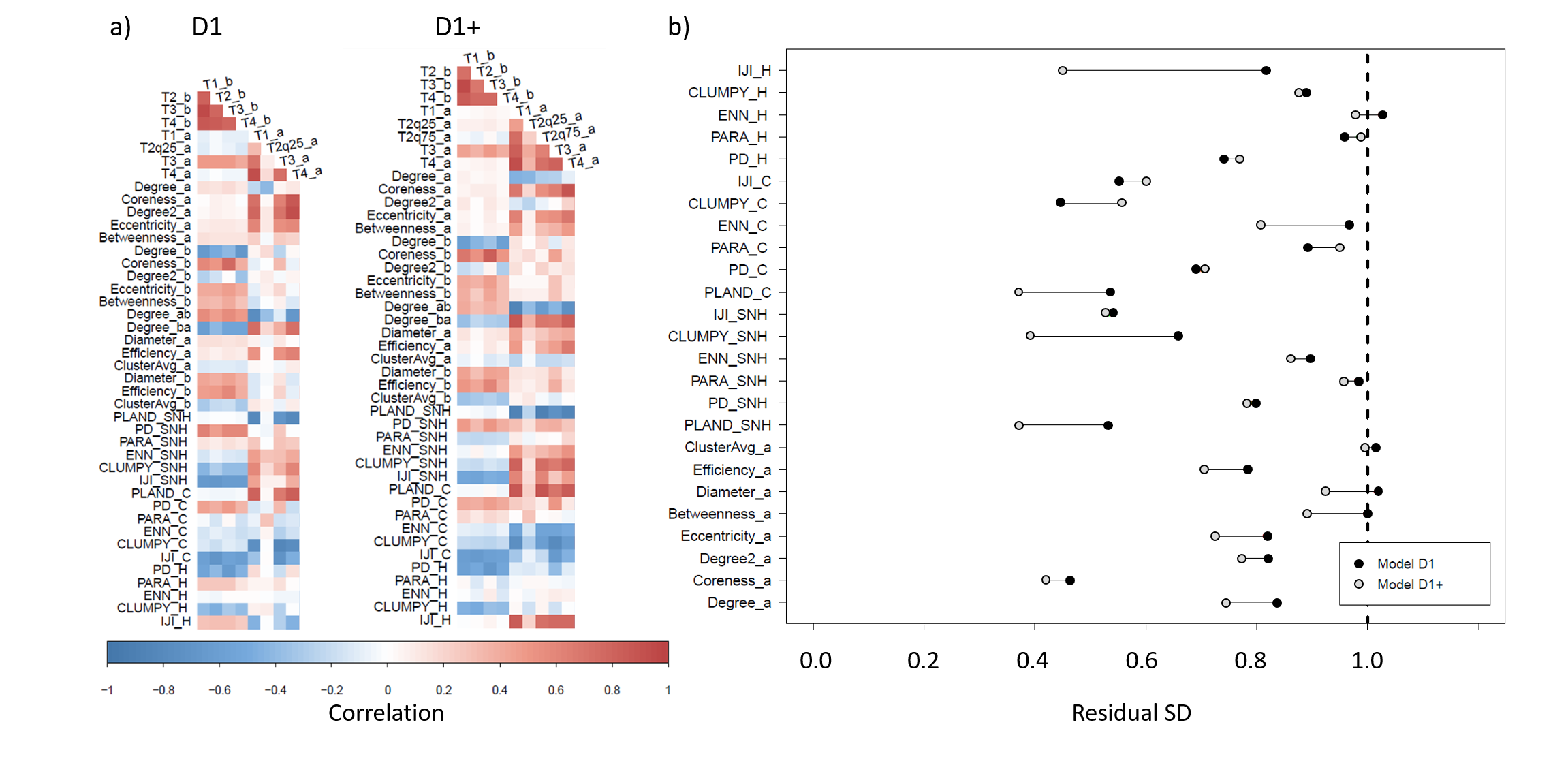} 
\caption{Correlation analysis. Panel a) Correlations among landscape descriptors and metrics in models D1 and D1+. For network metrics, letters stand for the network layer: $a$ = patch, $b$ = linear element, $ab$ = patch-linear element interaction. In raster metrics, SNH stands for Semi-natural habitat, C stands for Crop and H stands for hedges.
Panel b) Evolution of patch-related metrics from D1 do D1+ based on standard deviation not explained by the model's landscape descriptors.}
\label{fig:corr_SD}
\end{figure}

Different landscape descriptors and metrics may comprise similar information, and then non negligible correlation arises among such variables. If we seek a realistic representation of a specific metric through the model, then the landscape descriptors included in the model (or combinations of them) should be correlated with this metric. To assess this relationship, we use linear regression with the landscape descriptors as predictors and landscape metrics as dependent variables, and then consider the part of the standard deviation of the response not explained by the predictors.  Including additional descriptors can then improve the performance with respect to network- or grid-based metrics of interest. For illustration, we analyse the difference among the models D1 and D1+ through the correlation analysis in Figure \ref{fig:corr_SD}a. The descriptor formulated with respect to  the $25\%$-quantile of patch sizes  is relatively uncorrelated to other descriptors and metrics for $D1$. However, the $75\%$-quantile descriptor additionally  included in  $D1+$ is often highly correlated with other metrics for crop patches. In cases with strong negative or positive correlation, this tends to improve the validation performance in network-scale and raster metrics, see the diagram of Figure \ref{fig:corr_SD}. It shows the evolution in the residual standard deviations not explained by the descriptors of the models. Moreover, Monte-Carlo pseudo-$p$-values in Table~\ref{tab:MC_test} improve strongly from $D1$ to $D1+$. Detailed results for the three study domains provided in the Supplementary Material
indicate generally good performance for  the \textit{large region D3}.

\begin{table} 
\centering
\caption{\label{tab:MC_test} Pseudo-$p$-values of network-scale metrics and raster-based metrics for  the \textit{small region D1} with the basic model (D1) and the extended one (D1+).}
\begin{tabular}{|l|cc|cc|cc|}
\hline
 & \multicolumn{2}{c|}{\textbf{Semi-natural}} & \multicolumn{2}{c|}{\textbf{Crop}} & \multicolumn{2}{c|}{\textbf{Hedge}} \\ \cline{2-7} 
 & D1 & D1+ & D1 & D1+ & D1 & D1+ \\ \hline
\emph{Diameter} & - & - & 0.26 & 0.57 & 0 & 0.19 \\
\emph{Efficiency} & - & - & 0.06 & 0.56 & 0 & 0.23 \\
\emph{Cluster average} & - & - & 0.13 & 0.16 & 0.01 & 0 \\ \hline
PLAND & 0 & 0.08 & 0 & 0.10 & 0.03 & 0 \\
PD & 0 & 0 & 0.26 & 0.44 & 0.41 & 0.37 \\
PARA\_MN & 0.20 & 0.20 & 0.33 & 0.12 & 0.04 & 0.6 \\
ENN\_MN & 0.11 & 0.49 & 0.24 & 0.30 & 0.01 & 0.01 \\
IJI & 0.02 & 0.45 & 0.19 & 0.47 & 0.08 & 0.47 \\
CLUMPY & 0.11 & 0.19 & 0.02 & 0.24 & 0 & 0 \\ \hline
\end{tabular}
\end{table}

\section{Conclusion} \label{Concl}

Simulation of variable virtual landscapes opens up new ways of exploring various environmental issues \citep{Langhammer2019}. We have developed stochastic agricultural landscape models and statistical inference with a focus on the land-use allocation mechanism of patches and linear elements, using network models as an intuitive and flexible tool allowing for direct control and interpretation with respect to local landscape behavior. We have focused on descriptors based on single objects or pairs of such, leading to a certain robustness of modeling, estimation and simulation procedures. The sufficient summary statistics in the exponential family models in the  data application to the Lower Durance Valley were satisfactorily reproduced by simulations from fitted models, especially after inserting a new descriptor in the extended model D1$+$. We conclude that the fit is good, such that the model succeeds in capturing the key patterns of configuration and composition in real landscapes. 

From a functional point of view, vector-based models such as ours are more parsimonious and meaningful \citep{Gaucherel2012, Bonhomme2017}, and they enable handling different spatial and temporal scales. In pixel-based approaches, an appropriate representation of small-surface elements such as hedges would require a very high raster resolution, while a homogeneous large-surface patch would be made up of a very large number of pixels, instead of a single geometric object in our model. 
Our multiplex network structure assures low computational cost and memory requirements.

We have checked the generality of calibrated models by studying how simulated landscape metrics  are correlated to observed ones, using metrics that are not explicitly encoded into the model structure.  Certain data patterns calculated through metrics on raster scale are not correctly reproduced by our models, but this is partly explained by  instabilities in treating small-scale small-area patterns that are inherent to raster discretizations. 
Linear element allocation also shows some discrepancy between model and data for  large-scale clustering properties. To remedy the issue of appropriately reproducing an important landscape descriptor that is not directly controlled by the model, we can include related descriptors into the model \citep{Kleijnen1995}, or add additional constraints during simulation using techniques such as Simulated Annealing. 

We point out the potential of Approximate Bayesian Computation \citep[ABC, e.g.][]{Grelaud2009} for parameter estimation and for the \emph{inverse problem} of finding the model parameters that allow generating specific target values of landscape descriptors. Estimation through ABC is possible by setting target values equal to  observed landscape descriptors, and is asymptotically consistent under mild conditions if the set of target descriptors corresponds to the descriptors  $T_k$  in the Gibbs energy \eqref{eq:efm}. ABC can be useful for likelihood-free model selection using Bayes factors \citep{Grelaud2009}. However, unreported preliminary results show that rather long computation times arise with this method. 

We do not directly model human action  in the temporal dynamics of agricultural environments \citep{Bonhomme2017, Poggi2018}. For this, we would have to couple our model with a decision tool. Future developments also comprise the integration of our  land-use allocation model  with (existing) generative parametric tessellation model for the geometrical support.

\section*{Acknowledgements}
We are thankful to Claire Lavigne and Katarzyna Adamczyk for their help and wise suggestions for landscape data processing and for the discussion part.


\bibliographystyle{apalike}
\bibliography{References}

\end{document}